\documentclass[letterpaper,10pt,conference]{ieeeconf}
\IEEEoverridecommandlockouts
\usepackage[english]{babel}
\usepackage[utf8x]{inputenc}
\usepackage[T1]{fontenc}

\usepackage{amsmath,cite}
\usepackage{graphicx}
\usepackage[colorinlistoftodos]{todonotes}
\usepackage[colorlinks=true, allcolors=blue]{hyperref}

\title{\LARGE \bf
Semi-Interpenetrating Cooperative Localization in  Connected Vehicle Networks}
\author{Macheng Shen$^{1}$, Huajing Zhao$^{2}$, Jing Sun$^{3}$, Ding Zhao$^{2}$%
\thanks{$^{1}$M. Shen is with the Mechanical Engineering at the Massachusetts Institute of Technology, Cambridge, MA, USA, 02139.}%
\thanks{$^{2}$H. Zhao and D. Zhao ({\tt\small corresponding author: zhaoding@umich.edu}) are with the Department of Mechanical Engineering, University of Michigan, Ann Arbor, MI, 48109 USA.}%
\thanks{$^{3}$J. Sun is with the Department of Naval Architecture and Marine Engineering, University of Michigan, Ann Arbor,
MI, 48109.}%
}

\begin{document}
\maketitle
\begin{abstract}

We proposed a fusion mechanism for the distributed cooperative map matching (CMM) within the vehicular ad-hoc network. This mechanism makes the information from each node reachable within the network by other nodes without direct communication, thus improving the overall localization accuracy and robustness. Each node runs a Rao-Blackwellized particle filter (RBPF) that processes the Global Navigation Satellite System (GNSS) measurements of its own and its neighbors, followed by a map matching step that reduces or eliminates the GNSS atmospheric error. Then each node fuses its own filtered results with those from its neighbors for a better estimation. In this work, the complicated dynamics and fusion mechanics of these RBPFs are represented by a linear dynamical system. We proposed a distributed optimization framework that explores the model to improve both robustness and accuracy of the distributed CMM. The effectiveness of this distributed optimization framework is illustrated by simulation results on realistic vehicular networks drawn from data, compared with the centralized one and a decentralized one with random fusion weights.
\end{abstract}

\section{Introduction}

Global Navigation Satellite System (GNSS) is a navigation system that uses satellites to provide geo-spatial positioning with global coverage, which allows electronic receiver to calculate its current location with high precision and time synchronization. Low-cost GNSS are used for mobile applications in most cases with localization error up to several meters. The limited accuracy of the low-cost GNSS is mainly due to the atmospheric effects, referred to as the correlated error, as well as receiver noise and multipath offsets, referred to as the non-correlated error. Various techniques for reducing GNSS localization error have been investigated, such as precise point positioning (PPP) \cite{hofmann2012global}, real time kinematics (RTK) \cite{wang1999stochastic} and sensor fusion \cite{sasiadek1999sensor}.
However, few have been designed to reach the desired fusion precision without incurring additional hardware and infrastructure costs. Therefore, cost effective solutions for lane level accuracy are highly desired, especially for the autonomous vehicle applications in safety critical situations.

Recent research activities have been motivated by the demand for enhancing the localization accuracy of GNSS without rising implementation costs. These efforts have led to several noticeable developments, among which cooperative GNSS localization \cite{obst2012cooperative} and cooperative map matching (CMM) \cite{rohani2016novel,shen2017optimization,ramachandran2007experimental} are the most distinctive ones. 
With the Dedicated Short Range Communication (DSRC), cooperative localization that combines the group of GNSS measurements for high accuracy has been investigated by researchers, e.g., \cite{IEEE_magazine,rife, rohani2016novel, shen2017improving}. 
Additional measurements provided by the vehicle network lead to new constrains that CMM can exploit to reduce common errors, thereby improving localization performance.
\begin{figure}[t!]
  \centering
  \includegraphics[width=0.9\columnwidth]{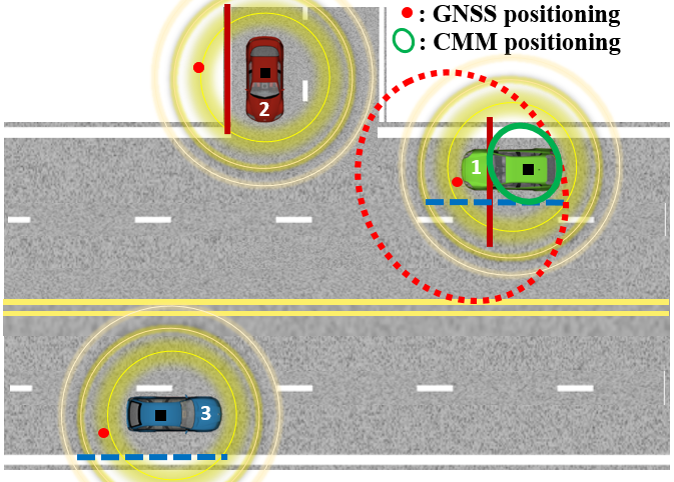}   
  \caption{Illustration of CMM involving three vehicles}
  \label{illustration}
\end{figure}
The key element in CMM is to combine informations collected by individual vehicles. Several fusion mechanisms for the distributed CMM within the vehicular ad-hoc network have been developed aiming at improving localization accuracy by matching the GNSS positioning of a group of connected vehicles to a digital map. In this work, we provided a model to represent the complicated dynamics and fusion of Rao-Blackwellized particle filters (RBPFs) as a linear dynamical system. A distributed optimization framework was proposed to exploit this model in an attempt to enhance both robustness and accuracy of the distributed CMM. Through similarities of a realistic vehicle networks, we showed that CMM can achieve desired localization accuracy while maintaining low-cost.  

The remaining of this paper is organized as follows: In Section II, we introduce the motivations on developing decentralized CMM methods as well as the reasons for using semi-interpenetrating cooperative vehicle networks. In Section III, we introduce the notations, the mathematical modeling and the formulation of the decentralized CMM with fusion mechanism. In Section IV, we present the simulation results of the centralized CMM, the decentralized CMM with optimized fusion, as well as a decentralized CMM with random fusion on three networks with different vehicle penetration rates. In Section V, we summarize the results and contributions of this work.

\section{Motivation}
\indent Our prior work \cite{shen2017improving} has developed an RBPF that filters the uncorrelated error and eliminates the effects of correlated error by cooperative map matching (CMM). CMM, first proposed in Rohani et al. \cite{rohani2016novel}, is a method that matches the GNSS positioning of a group of vehicles to a digital map and corrects the biased positioning by enforcing the road constraints. The effectiveness of CMM depends on the diversity of the road constraints. Fig. \ref{illustration} shows the raw GNSS positioning (red dot and red ellipsoid) and the corrected one by CMM (green ellipsoid), and the effects of road constraints are demonstrated. In prior work \cite{shen2017impact}, we quantified the correlation between the diversity of the road constraints and the CMM localization error, and developed algorithms that optimally design the connection network subject to limited communication bandwidth \cite{shen2017optimization}. 
\subsection{Why Decentralized CMM?}
\indent The methods developed \cite{shen2017impact,shen2017improving} implicitly assumed a centralized architecture that gathers all the GNSS measurements for fusion and optimization. In real vehicular networks, especially the large ones, a decentralized architecture is more desirable and realistic for the following reasons:
\begin{enumerate}
 \item The communication distance of DSRC is limited. Vehicles far away from each other cannot directly communicate.
 \item The communication bandwidth is limited. Each vehicle can only communicate with up to 30 other vehicles without significant package loss \cite{ramachandran2007experimental}.
 \item The on-board computational capacity is limited. As the computational complexity grows with the number of vehicles involved, on-line filtering of raw data could become challenging. 
 \end{enumerate}

\indent One straightforward way to overcome the communication constraints is to implement individual RBPF on each vehicle using the locally available GNSS measurements from itself and all of its neighbors, to update the estimation. Unlike centralized CMM, decentralized CMM networks requires less computation capacity, and is thus a more realistic structure for real-life connective vehicular networks.

\subsection{Semi-Interpenetrating Vehicular Networks}
It is expected that the reduced number of vehicles involved in commutation would lead to a sparse network where the number of links connecting nodes are sparse. This could in turn lead to large localization error, as there would not be enough road constraints to mitigate the common GNSS error. In view of this limitation, decentralized CMM along with Semi-Interpenetrating Cooperation Networks is a desired solution for balancing the communication capacity and the localization accuracy.
 
In Semi-interpenetrating vehicle networks, each vehicle not only use the measurements from neighbors to update their own RBPF estimations, but also fuse the RBPF estimations of neighbors. This provides an innovative method to partially information information from nodes that are not directly connected to the receiver without expanding communication capacity. As the estimations are represented by RBPFs, the fusion means simply stacking the particles and eliminating those inconsistent with the road constraints. Information from any node can be propagated to any other node by repeated local fusion as long as the graph representing the network is connected. We also expect that the resulted average localization error would be much smaller than that of the decentralized CMM without fusion. Furthermore, the localization error can be minimized if the fusion is designed to optimize certain error criterion.
\section{Mathematical detail}
In this section, we introduce the notations used to describe the vehicular network, the mathematical modeling of the RBPF dynamics and the decentralized optimization of the fusion mechanism. 
\subsection{Notations}
We first define the connected vehicle network as a undirected graph $(V,E)\in G$, where $V=\{1,2,3,...N\}$ is the set of nodes each represents a vehicle within the network and $E$ is the set of edges within the network. If two vehicles $i$ and $j$ are within communication ranges, then there is an edge $(i,j)\in E$ between these two nodes. Nonetheless, it does not imply that there has to be a communication between these two vehicles. We use $x_i$ to denote the common error estimated by vehicle $i$, which is calculated by averaging all the common error estimations, represented by the particles belonging to vehicle $i$.


\subsection{Modeling RBPF dynamics and fusion}
The RBPF uses a predict-update framework to estimate the common error. In the prediction step, vehicle positions common error are predicted by each particle with random Gaussian noise added. 
In the update step, those particles incompatible with the road constraints have a large probability of being eliminated. As a result, the RBPF estimation after map matching is biased towards the direction where the particles obey the road constraints. If each vehicle only has access to its own GPS or only few other vehicles' GPS, then it is likely that the RBPF would diverge due to the random Gaussian noise added in the prediction step. That is, the estimated common error would deviate from the ground truth without bound. This type of divergence happens when the road constraints of the communicated vehicles are not sufficiently diverse.\\
\indent The divergence in the common error estimation can be mitigated by richer information of road constraints through repeated fusions of the estimated common error from neighboring nodes. 
As both the measured position and the predicted position of each vehicle would be passed to its neighboring node,
hypothetically, if the fusion is conducted N times, where N is equal to or larger than the diameter of the network
(the largest path length between any two nodes), then each node implicitly utilizes the road constraints "seen" of all the nodes of the network, as these constrains are embedded in the RBPFs of the nodes.\\
\indent We, therefore, propose a fusion mechanism for the CMM vehicular network using RBPF described as follows:
\begin{enumerate}
 \item For each vehicle within the network, take particles from their neighboring vehicles and stack them together with its own particles.
 \item Calculate weights according to the map matching and resample to downsize the number of the particles to the original size.
 \item Obtain the numbers of particles taken from neighboring vehicles by optimizing certain criterion.
 \end{enumerate}
\indent We model the evolution of the networked RBPF estimations as a coupled linear system, described by the following equation:
\begin{equation}
x_i(t+1)=\sum\limits_{j=1}^N a_{i,j}(t)x_j(t)+w_i(t),
\label{1}
\end{equation}
where
\begin{equation}
a_{i,j}(t)=0,\forall (i,j)\notin E.
\end{equation}
\indent Here, $x_i(t)$ denotes the estimated common error of vehicle $i$ at timestep $t$, while $w_i(t)$ denotes a random Gaussian noise applied to $i$ at the same timestep.
Eq. \ref{1} means that the estimated common error after fusion equals to a weighted average over the neighboring estimation and the own estimation at the previous time instance, added by some random noise from the RBPF prediction. This equation can explain the reason why the RBPF diverges if none of the vehicles fuses estimations from other vehicles, simply by taking $a_{i,j}=0,i\neq j; a_{i,i}=1$. This represents a discrete random walk that is driven by the random force $w_i(t)$ and can be unbounded. We expect that this linear modeling of the particle filter dynamics is an approximation to the true dynamics when all the particles are reasonably close to the true common error such that all the particles receive weights of comparable magnitude. As a result, the resampling process described in the fusion mechanism is like a noisy weighted averaging over the particle estimations. 
\subsection{Minimize variance}
\indent Eq. \ref{1} leaves freedom for us to select the fusion coefficients $a_{i,j}$ by selecting the different numbers of particles taken from neighboring vehicles. It is desirable to select the numbers of particles such that the averaged CMM localization error over the whole network is minimized. That is, we want to choose $a_{i,j}$ subject to the communication constraints such that we can minimize 
\begin{equation}
\begin{aligned}
J&=\frac{1}{N}\sum\limits_{i=1}^N (x_i(t+1)-c(t+1))^2\\
&=\frac{1}{N}\sum\limits_{i=1}^N(x_i(t+1)-\bar{x}(t+1))^2\\
&+\sum\limits_{i=1}^N(\bar{x}(t+1)-c(t+1))^2,
\end{aligned}
\label{error_decom}
\end{equation}
where $c(t+1)$ is the true common error at the time instance $t+1$, which is assumed to be slowly time-varying; $\bar{x}(t+1)=\frac{1}{N}\sum\limits_{i=1}^N x_i(t+1)$ is the mean of the common error estimation over the whole network.\\
\indent Eq. (\ref{error_decom}) has a clear interpretation: The objective $J$ represents the averaged squared estimation error over the network. It can be decomposed into two parts: The variance of the estimations over the network and the squared error of the mean estimation over the network. However, it is impossible to directly minimize the objective $J$, as the true common error is unknown. We, thereby, aim to minimize the variance in hope that by doing so, the original objective function $J$ remains small. We present the following arguments for this choice:
\begin{itemize}
\item The original objective $J$ is always larger of equal to the variance. It is necessary to keep the variance small in order to keep $J$ small.
\item The variance can be calculated given all the estimations within the network in a decentralized manner, with $\bar{x}$ obtained by distributed consensus algorithms.
\item As the variance becomes small, all the estimations within the network tend to the same estimation, which could lead to a small estimation error of the mean as a result of CMM. The original objective $J$ would also be small.
\end{itemize}
\indent Given the aforementioned rationals, we formulate the following optimization problem:
\begin{equation}
\begin{aligned}
minimize\mbox{}&\tilde{J}=\frac{1}{N}\sum\limits_{i=1}^N(x_i(t+1)-\bar{x}(t+1))^2,\\
subject\mbox{ }to\mbox{ }&x_i(t+1)=\sum\limits_{j=1}^N a_{i,j}(t)x_j(t),\\
&a_{i,j}(t)=0,\forall (i,j)\notin E,\\
and\mbox{ }& 0\leq a_{i,j}(t)\leq 1.
\end{aligned}
\label{opt_problem}
\end{equation}
\indent The fusion coefficients $a_{i,j}(t)$ are the free variables to be selected. Note that the objective function is the sum of all local objective functions, Eq. (\ref{opt_problem}) is essentially a decentralized quadratic programming.
\subsection{Accelerate consensus}
In Section III.C, we show that it is necessary to bound the variance in order to bound the mean squared estimation error. The intuition is that as the disagreement between the nodal estimations becomes small, the estimation error tend to be bounded by map matching. Besides variance, the convergence rate of Eq. (\ref{1}) is another metric that measures the agreement and disagreement of node decisions within a network. It is expected that by maximizing the convergence rate, the mean squared localization error would also be small. It has been well studied how to maximize the convergence rate given a network with fixed structure in Xiao \cite{xiao2004fast}. We briefly mention the major results here:\\
Given a linear consensus iteration:
\begin{equation}	
X(t+1)=AX(t),
\end{equation}
where
\begin{equation}
A_{i,j}=0,\forall (i,j)\notin E.
\end{equation}
The asymptotic convergence rate is defined as:
\begin{equation}
r_{asym}(A)=\mathop{sup}\limits_{X(0)\neq \bar{X}}\lim_{t \to \infty}(\frac{||X(t)-\bar{X}||_2}{||X(0)-\bar{X}||_2})^{\frac{1}{t}},
\end{equation}
where $\bar{X}$ is the fixed point of this linear iteration.\\
\indent There is a systematic approach to maximize $r_{asym}$ by solving an optimization problem that requires the knowledge of the network topology. This would be problematic for all problem because of the communication limit. An alternative approach that does not give optimal convergence rate but only uses local topology is given below \cite{xiao2004fast}:
\begin{equation}
A_{i,j}=\frac{1}{max\{d_i,d_j\}},
\label{max_degree}
\end{equation}
where $d_i$ is the degree of the node $i$, which is equal to the number of neighbors of the node $i$. This approach guarantees convergence as well as reasonably fast convergence rate.
\section{Simulation results}
In this section, we present simulation results to illustrate the performance of the proposed fusion mechanism.
\subsection{Simple network involving four vehicles}
We first present the simulation results on a simple network involving four vehicles at a cross road to demonstrate that model. Each vehicle can only communicate to itself and one another vehicle. The network connection matrix is as follows:
\begin{equation}
E_C=\left[
\begin{matrix}
1&1&0&0\\
0&1&1&0\\
0&0&1&1\\
1&0&0&1\\
\end{matrix}
\right],
\end{equation}
where $E_C(i,j)=1$ means that vehicle $i$ can receive signals from vehicle $j$.\\
\indent We compare the following two approaches:\\
(1) Weight determined by the variance minimization presented in Section III.C.\\
(2) Constant weight matrix:
\begin{equation}
A=\left[
\begin{matrix}
\alpha&1-\alpha&0&0\\
0&\alpha&1-\alpha&0\\
0&0&\alpha&1-\alpha\\
1-\alpha&0&0&\alpha\\
\end{matrix}
\right],
\end{equation}
where $\alpha=0.5$ would correspond to Eq. (\ref{max_degree}).\\
\begin{table}[ht] 
\caption{Mean Squared Error (MSE) and Variance} 
\centering      
\begin{tabular}{c c c}  
\hline\hline                        
Approach&$\sqrt{Variance}$ (m)&$\sqrt{MSE}$ (m)\\ [0.3ex] 
\hline                    
Variance Minimization&0.13&0.58\\ 
Constant $\alpha=0.05$&0.38&0.81\\ [0.3ex]
Constant $\alpha=0.1$&0.33&0.63\\
Constant $\alpha=0.2$&0.28&0.6\\
Constant $\alpha=0.4$&0.21&0.56\\
Constant $\alpha=0.5$&0.2&0.57\\
Constant $\alpha=0.6$&0.21&0.58\\
Constant $\alpha=0.8$&0.33&0.59\\
Constant $\alpha=0.9$&0.46&0.62\\
Constant $\alpha=0.95$&0.68&0.8\\
\hline     
\end{tabular}
\label{var_err}
\end{table}
Table \ref{var_err} shows the simulated results, which reveals a correlation between the variance and the mean squared error (MSE). In general, as the variance decreases, so does the MSE. Nonetheless, this correlation is not a deterministic relationship. While the variance becomes small enough, the correlation becomes weak. This can be explained by Eq. (\ref{error_decom}). As the variance becomes small, the other term becomes dominant, thus weakening the correlation between the MSE and the variance. Nevertheless, the results provide the possibility to minimize variance as a substitution of error
minimization.
\subsection{Large network involving fifty vehicles}
\begin{figure}[htbp]
  \centering
  \includegraphics[width=0.9\columnwidth]{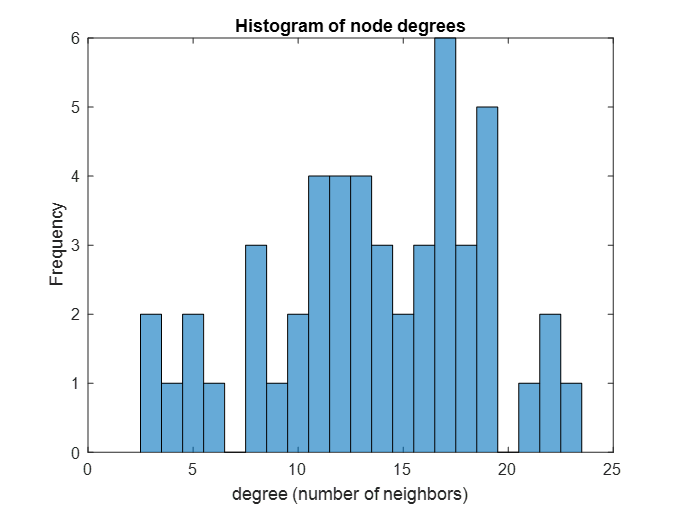}   
  \caption{Node degree distribution corresponding to the network $\mathcal{N}_1$}
  \label{dense_degree}
\end{figure}

\begin{figure}[htbp]
  \centering
  \includegraphics[width=0.9\columnwidth]{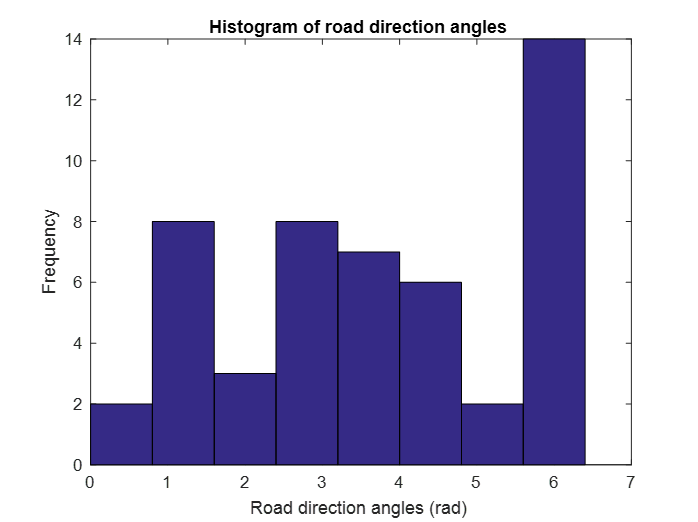}   
  \caption{Road direction angle distribution corresponding to the network $\mathcal{N}_1$}
  \label{road_angle}
\end{figure}

We present the 
MSE and variance of the estimations over the nodes
on a large network involving fifty vehicles. The locations and road direction angles of these vehicles are sampled from a joint distribution $p(x,\theta)$, which is estimated from the historic traffic data of three years period recorded in Ann Arbor.\\
\indent We first assume the the maximum communication radius is 3 kilometers. The connection matrix of this vehicular network can be obtained given the locations of the vehicle. The histogram of the nodal degrees (number of neighbor) and road direction angles are shown in Fig. (\ref{dense_degree}) and Fig. (\ref{road_angle}). We refer to this network as $\mathcal{N}_1$.\\
\begin{figure}[htbp]
  \centering
  \includegraphics[width=0.95\columnwidth]{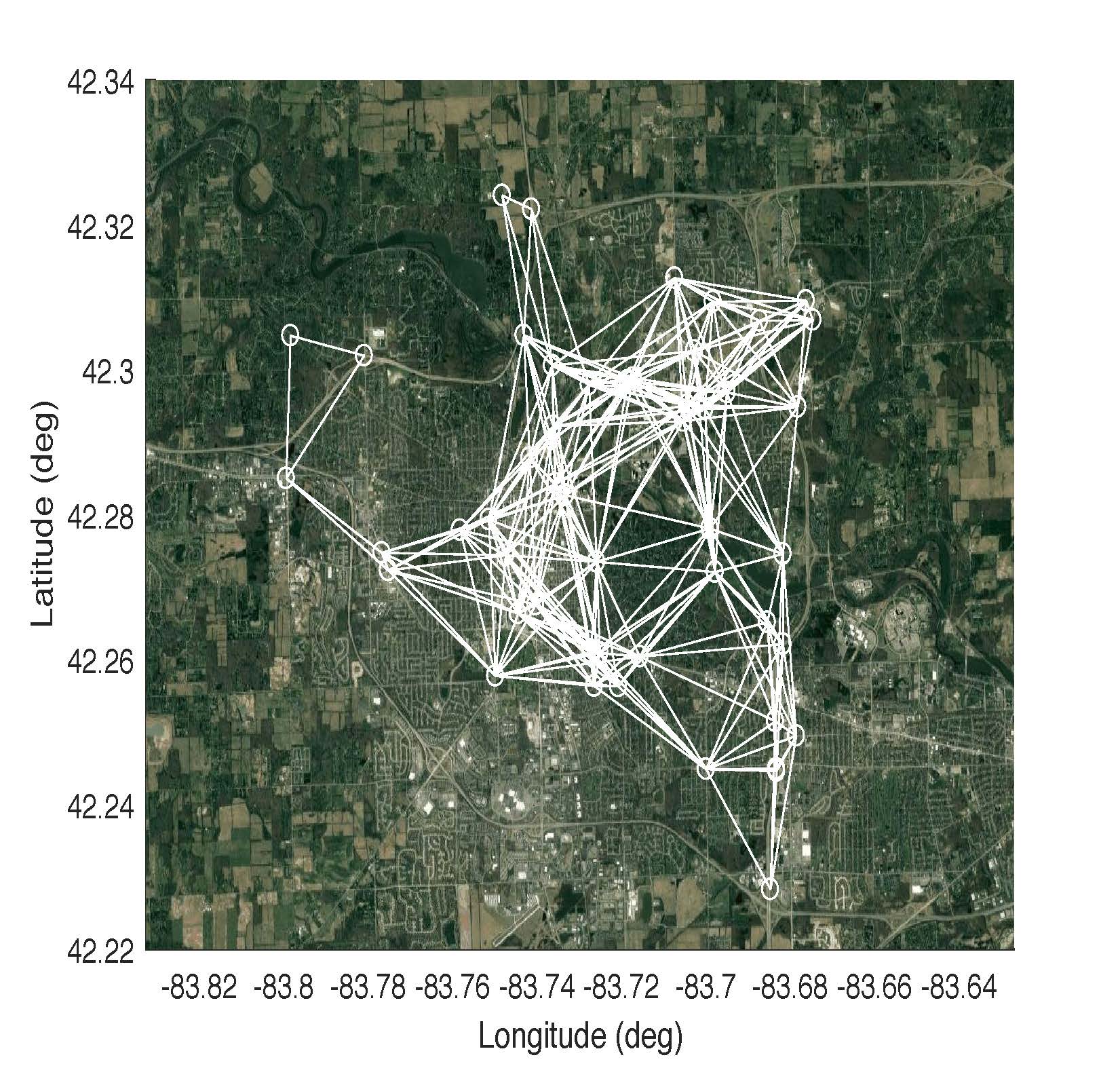}   
  \caption{Graph plot of the network $\mathcal{N}_1$}
  \label{net1}
\end{figure}
\begin{figure}[htbp]
  \centering
  \includegraphics[width=0.95\columnwidth]{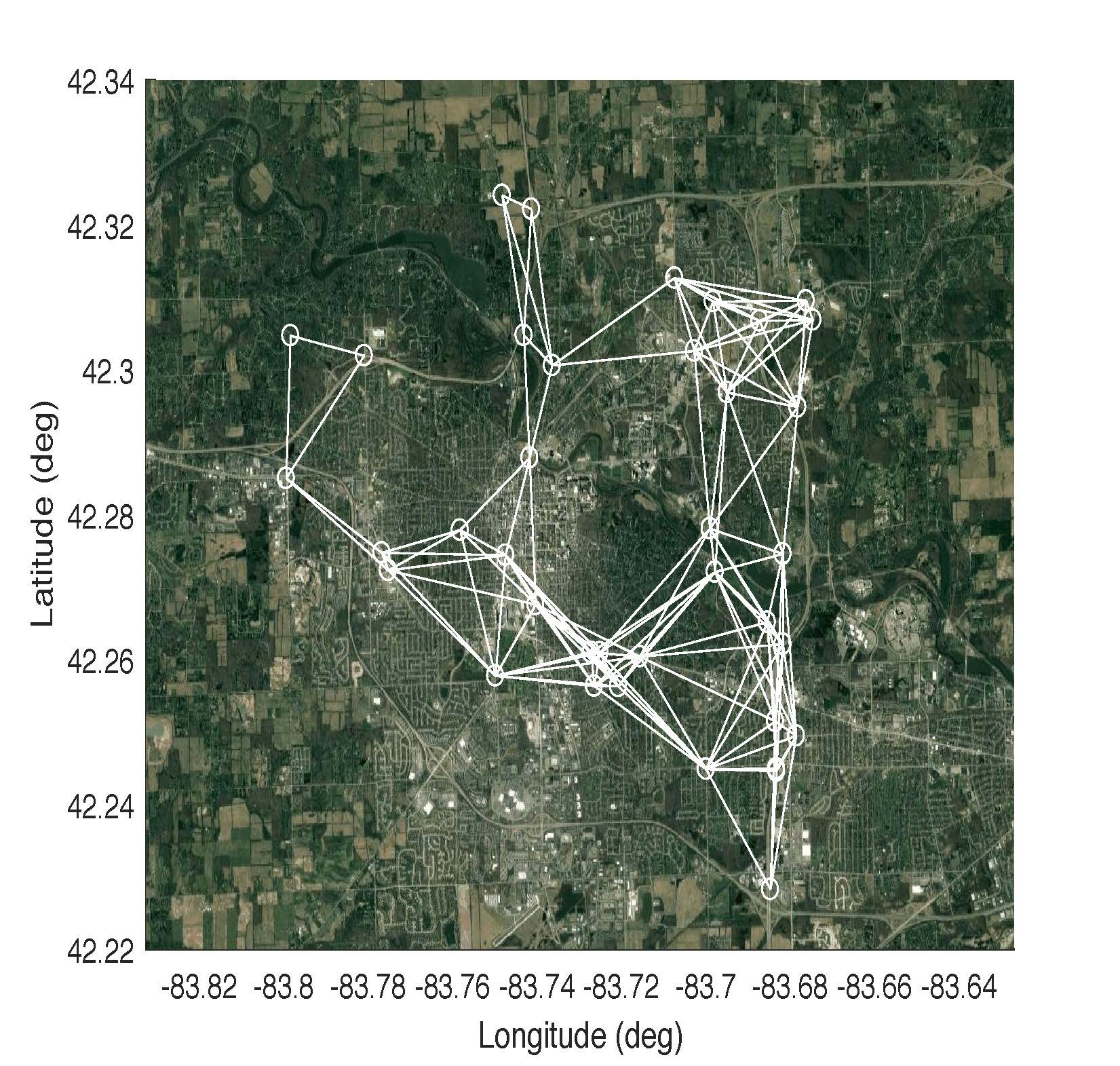}   
  \caption{Graph plot of the network $\mathcal{N}_2$}
  \label{net2}
\end{figure}
\begin{figure}[htbp]
  \centering
  \includegraphics[width=0.95\columnwidth]{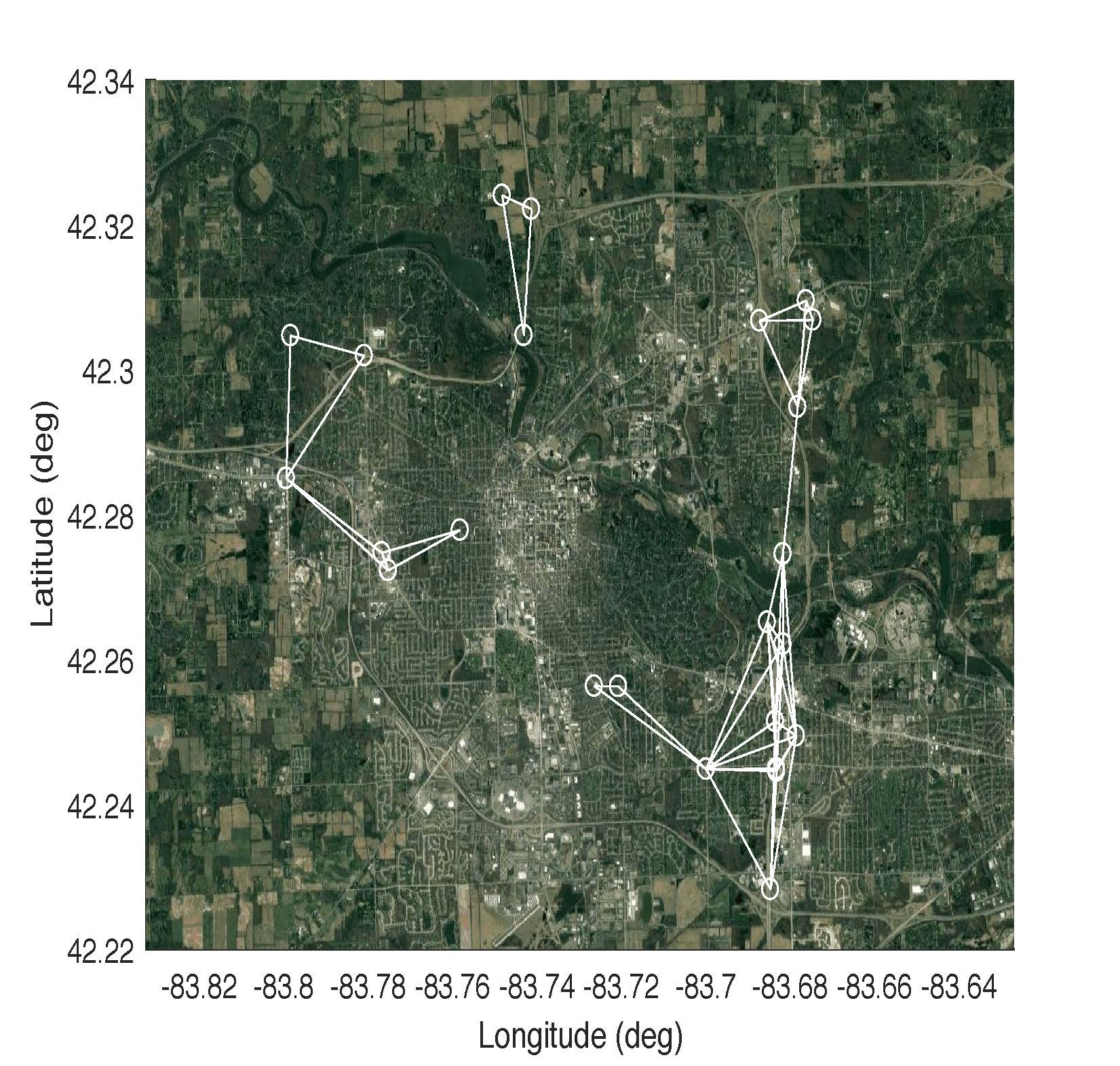}   
  \caption{Graph plot of the network $\mathcal{N}_3$}
  \label{net3}
\end{figure}
\indent Nonetheless, as this network is dense, we expect that the optimized fusion mechanism will not outperform much compared with random fusion, and that even the centralized CMM will not outperform much compared with the decentralized CMM. We, therefore, create two more networks by trimming edges from $\mathcal{N}_1$. More specifically, a sparser network $\mathcal{N}_2$ is created by deleting the nodes with degree larger than 20 from $\mathcal{N}_1$. Another even sparser network $\mathcal{N}_3$ is created in the same way by deleting the nodes with degree larger than 14 from $\mathcal{N}_1$. The remaining numbers of nodes in the network $\mathcal{N}_2$ and $\mathcal{N}_3$ are 38 and 24, respectively, which can be interpreted as approximately 75\% and 50\% penetration rate compared with the dense network. Graph plots of the network $\mathcal{N}_1$, $\mathcal{N}_2$ and $\mathcal{N}_2$ are shown in Fig. (\ref{net1}), (\ref{net2}) and (\ref{net3}), respectively.
We compare three different CMM mechanisms: 
\begin{itemize}
\item Centralized CMM: All the GPS measurements are gathered by a center node and processed by one single RBPF. This is regarded as the optimal localization results that can be achieved by CMM.
\item Decentralized optimized CMM: Each node processes its own and neighboring GPS measurements using a RBPF, and fuses its own estimation with its neighboring estimations in a weighted manner according to the optimization results from the optimization problem Eq. (\ref{opt_problem}).
\item Decentralized random CMM: Each node processes its own and neighboring GPS measurements using a RBPF, and fuses its own estimation with its neighboring estimations with a random weight.
\end{itemize}
\indent The root mean square error (RMSE) of these three CMM mechanisms on the three networks are listed in Table \ref{RMSE_large_net}.
\begin{table}[ht] 
\caption{Root Mean Squared Error (m)} 
\centering      
\begin{tabular}{c c c c}  
\hline\hline                        
CMM mechanism&Net $\mathcal{N}_1$&Net $\mathcal{N}_2$&Net $\mathcal{N}_3$\\ [0.3ex] 
\hline                    
Centralized&0.67&0.64&0.64\\ 
Decentralized Optimized&0.76&0.72&1.04\\
Decentralized Random&0.72&1.60&4.30\\
\hline     
\end{tabular}
\label{RMSE_large_net}
\end{table}
\begin{figure}[htbp]
  \centering
  \includegraphics[width=0.95\columnwidth]{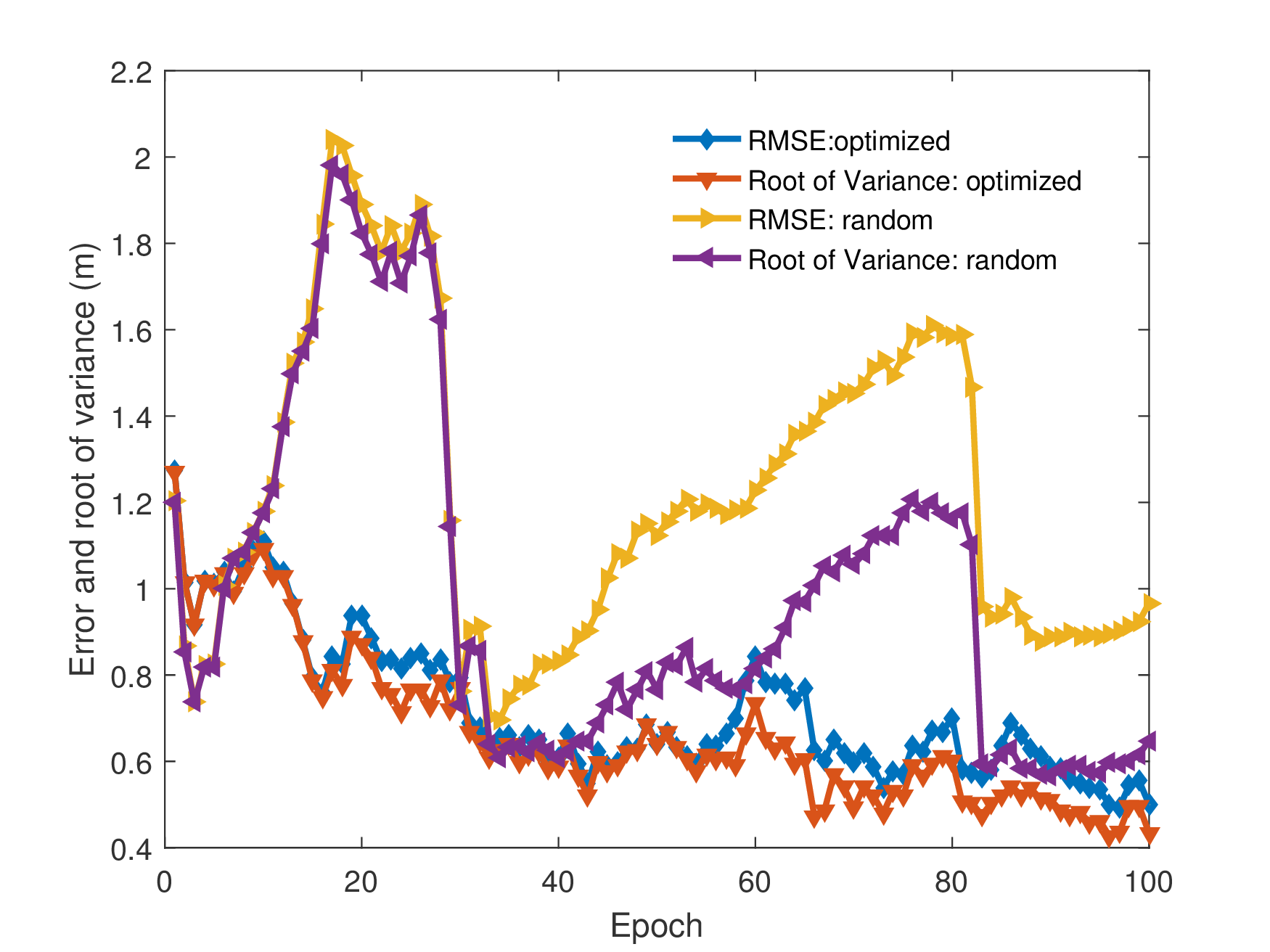}   
  \caption{RMSE and variance evolution on the network $\mathcal{N}_2$ using optimized decentralized CMM and random decentralized CMM}
  \vspace{-0.3cm}
  \label{net2_error}
\end{figure}
\begin{figure}[htbp]
  \centering
  \includegraphics[width=0.95\columnwidth]{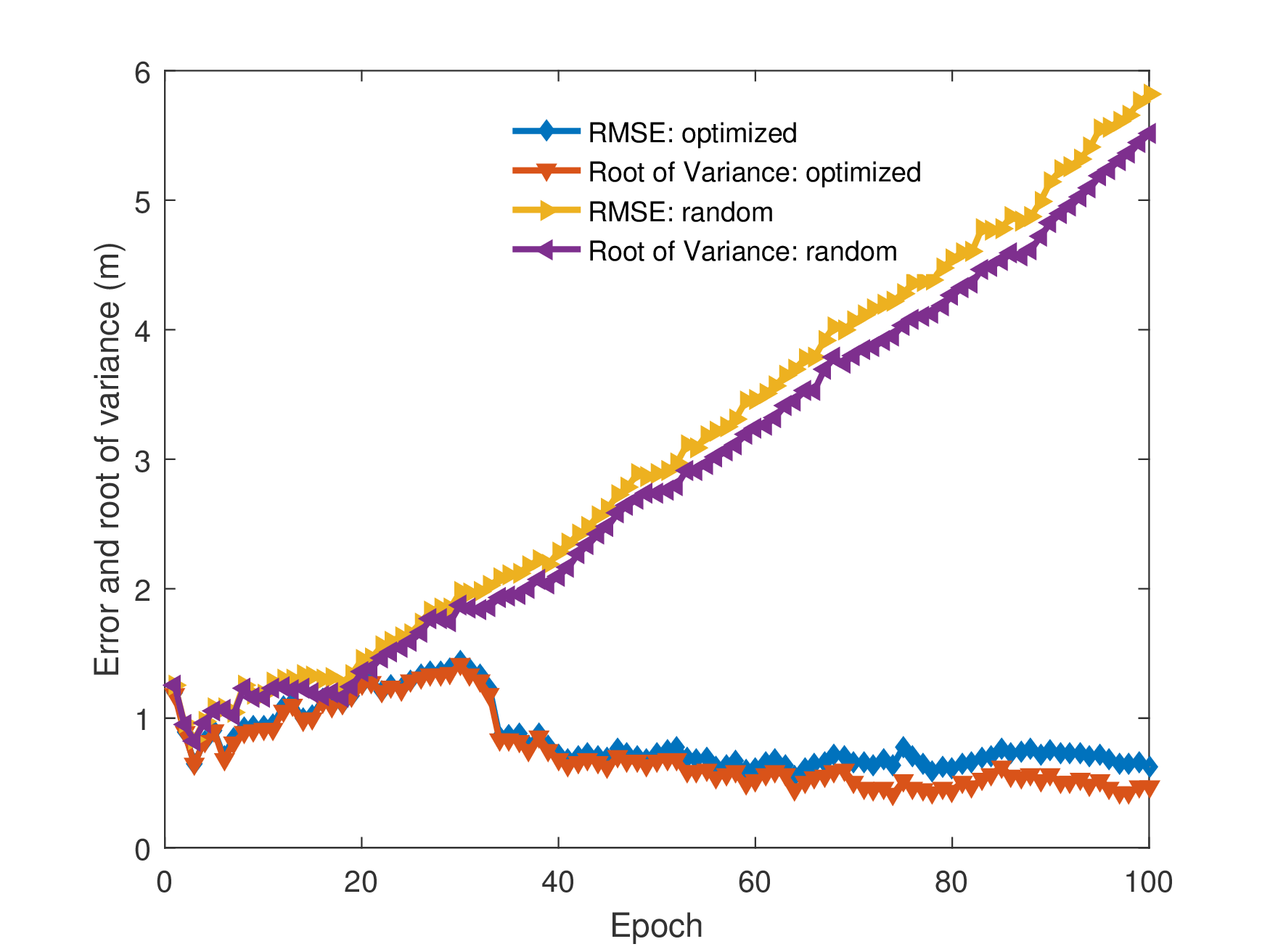}   
  \caption{RMSE and variance evolution on the network $\mathcal{N}_3$ using optimized decentralized CMM and random decentralized CMM}
  \vspace{-0.3cm}
  \label{net3_error}
\end{figure}
\indent Each value in Table \ref{RMSE_large_net} is averaged over three simulations. As expected, the centralized CMM achieves the smallest error. On the dense network $\mathcal{N}_1$, the differences between the three CMM mechanisms are very small, as most of the vehicles have many neighboring GPS measurements available. They achieve high localization accuracy even without sharing estimation with neighboring vehicles. On the relatively sparse network $\mathcal{N}_2$, it is clear that the optimized decentralized CMM significantly outperforms the random decentralized CMM. We identified that the localization error of some of the nodes grows monotonically using the random decentralized CMM. This diverging behavior happens when none of the particles is close enough the ground truth. As a result, the average localization error over the network is large. On the sparsest network $\mathcal{N}_3$, the centralized CMM significantly outperforms the other two decentralized ones, and the optimized decentralized CMM significantly outperforms the random one. The similar diverging behavior is also observed and becomes more serious when using the random decentralized CMM.
\indent Fig. \ref{net2_error} and Fig. \ref{net3_error} show the evolution of the RMSE and the variance over all the nodes in network $\mathcal{N}_2$ and $\mathcal{N}_3$, respectively. The RMSE and the variance are highly correlated, which verifies our proposition of minimizing variance as a surrogate of error minimization. This correlation implies that minimizing the variance is not only necessary for the purpose of minimizing the RMSE, but also usually sufficient. The random decentralized CMM results in a large RMSE on the relatively sparse network $\mathcal{N}_2$ and a diverging RMSE on the sparser network $\mathcal{N}_3$. This observation indicates that the proposed optimization framework increases the robustness of decentralized CMM.
\section{Conclusion}
This work provides a fusion mechanism for distributed CMM and shows the correlation between the estimation variance and the MSE over the network. 
In order to keep the error small, it is necessary to keep the variance small. Based on this observation, we proposed an optimization approach to determine the fusion weights by solving a decentralized quadratic programming. Simulation results verified the correlation between the estimation variance and the MSE. The optimized distributed CMM is shown to have higher accuracy and robustness than a random distributed CMM, especially when the network is sparse. 

Three observations can be made regarding Semi-Interpenetrating CMM Network. First, a remarkable correlation between RMSE and variance was demonstrated, which validates the feasibility to minimize RMSE by minimizing the variance.
Second, optimized decentralized CMM outperforms random decentralized CMM, especially on sparse networks.
Finally, random decentralized CMM can be quite inaccurate and even unstable on sparse networks, which is shown in Fig. \ref{net2_error} and Fig. \ref{net3_error}.
However, given the usual traffic flows in city-roads and highways, Semi-Interpenetrating CMM Network shows the potential to provide cooperative localization with desired accuracy for connected vehicles as a low-cost GNSS method.

\bibliographystyle{IEEEtran}
\bibliography{sample}

\end{document}